\documentclass[letterpaper, 10 pt, conference]{ieeeconf}  

\usepackage{graphicx}
\usepackage{subfigure}
\usepackage[bottom]{footmisc}
\IEEEoverridecommandlockouts                              

\title{\LARGE \bf
Lung Segmentation and Nodule Detection in Computed Tomography Scan using a Convolutional Neural Network Trained Adversarially using Turing Test Loss
}

\author{ Rakshith Sathish$^{1}$, Rachana Sathish$^{1}$, Ramanathan Sethuraman$^{2}$ and Debdoot Sheet$^{1}$
\thanks{*This work supported through a research grant from Intel India Grand Challenge 2016 for Project MIRIAD}
\thanks{$^{1}$Rakshith Sathish, Rachana Sathish and Debdoot Sheet are with the Department of Electrical Engineering, Indian Institute of Technology Kharagpur, India-721302
        {\tt\scriptsize \{rsatish, debdoot\}@ee.iitkgp.ac.in, rachana.sathish@iitkgp.ac.in}}%
\thanks{$^{2}$R. Sethuraman is with Intel Technology India Pvt. Ltd. Bangalore, India }%
}

\begin{document}

\maketitle
\thispagestyle{empty}
\pagestyle{empty}

\begin{abstract}
Lung cancer is the most common form of cancer found worldwide with a high mortality rate. Early detection of pulmonary nodules by screening with a low-dose computed tomography (CT) scan is crucial for its effective clinical management. Nodules which are symptomatic of malignancy occupy about 0.0125 - 0.025\% of volume in a CT scan of a patient. Manual screening of all slices is a tedious task and presents a high risk of human errors. To tackle this problem we propose a computationally efficient two stage framework. In the first stage, a convolutional neural network (CNN) trained adversarially using Turing test loss segments the lung region. In the second stage, patches sampled from the segmented region are then classified to detect the presence of nodules. The proposed method is experimentally validated on the LUNA16 challenge dataset with a dice coefficient of $0.984\pm0.0007$ for 10-fold cross-validation. 

\end{abstract}

\section{Introduction}
Lung cancer is the most common form of cancer worldwide, accounting for about 2.1 million new cases and 1.8 million deaths annually \cite{world2009world}. Presence of nodules is considered to be one of the precursors for lung cancer. Pulmonary nodules are radiographically opaque and measure up to 30 millimeter in diameter. Low-dose CT based imaging is performed to detect the presence of nodules \cite{national2011reduced}. The chances of survival increases significantly when diagnosed at an early stage of the cancer, thus rendering the task of nodule detection a critical one. 

Manual screening of nodules in CT scans is a time consuming and stressful task which requires the expertise of an experienced radiologist. A standard CT volume has 200-400 slices while the nodules are present in only 3-5 slices. Radiologists have to manually screen the slices, identify the suspicious slices and later detect nodules. Manual screening is often plagued with false negative classifications, due to the very small size of nodules. Automation of this task can lessen the workload of manual screening and also reduce false negatives and improve cancer management outcomes through early diagnosis.

\begin{figure}[!t]
  \centering
   \includegraphics[width=\linewidth]{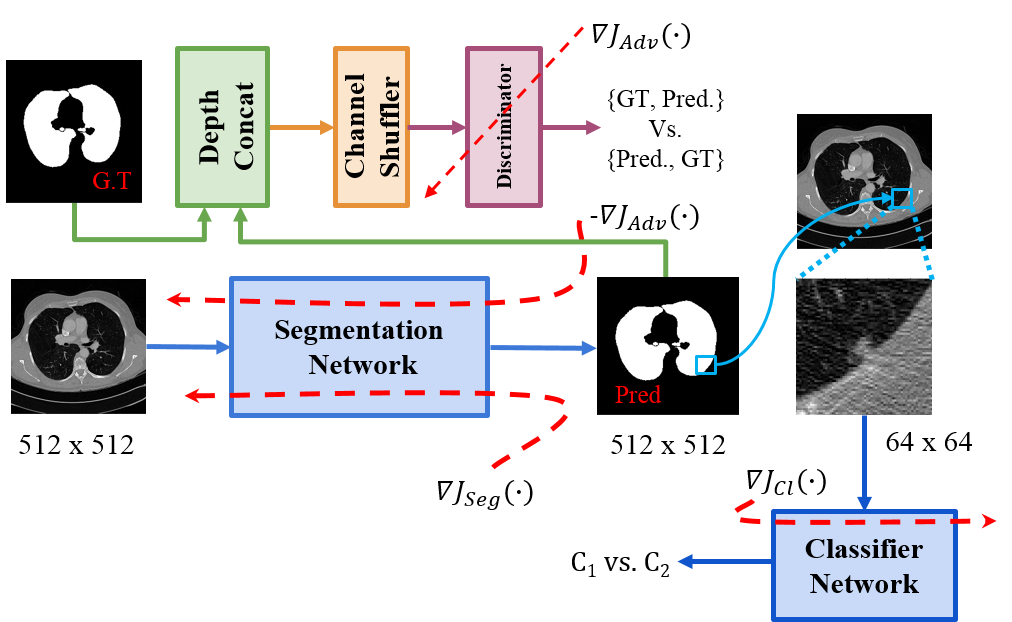}
  \caption{Overview of the proposed method}
  \label{fig:fig1}
\end{figure}

\textbf{Challenges:} One of the major challenges associated with the detection of lung nodules is the wide variation in appearance and spatial distribution. Nodules can be present either completely isolated within the lung parenchyma or attached to the lung walls. Characteristics of a nodule, such as texture, shape, opacity, size, etc have high variance which makes its detection a complex task.

\textbf{Related works:} Most of the existing methods propose to solve the problem of lung nodule detection through separate stages for lung and nodule segmentation. In one such method \cite{skourt2018lung}, lung segmentation has been performed using U-Net \cite{ronneberger2015u}. Recurrent residual convolutional neural  network (RRCNN) \cite{alom2018recurrent} have also been used for the same. An improved version of U-Net \cite{gu2019multi}, termed
multi-scale prediction network (MPN) was proposed to solve lung area segmentation problems on thoracic CT scans. A fast and efficient 3D lung segmentation method based on V-net was proposed by \cite{negahdar2018automated}. In \cite{fu2019automatic} the nodule detection task is performed in two stages. In the first stage, all possible candidate regions in CT slices are detected and in the later stage a 3D fully convolutional neural network (FCN) is used to detect nodules in the candidate regions with high certainty. Though this method presents high performance metrics, it also incurs high computational load in both stage for 3D processing.

\textbf{Our approach:} Considering the constraints and challenges associated with lung nodule detection, we propose a two stage framework as shown in Fig.~\ref{fig:fig1} involving only 2D convolutional neural networks (CNNs). In the first stage we implement a fully convolutional network trained adversarially to segment the lung region and in the second stage a classifier network is trained to identify presence of nodules in image patches extracted from within the segmented lung region. The proposed method involves 2D processing with a compute complexity of $157.320\times10^9$ FLOPS (floating point operations) and $2.796\times10^6$ FLOPS per slice for first and second stage respectively. 


  

\section{Problem Statement}
Consider a  CT volume $\mathbf{V}$ consisting of \textbf{$N$} number of axial 2D slice {$\mathbf{S}$}. We model the problem of nodule detection in two stages. In \emph{Stage 1}, each pixel of the slice $\mathbf{S_n}$ is classified into one of the two classes $\{lung, background\}$, to obtain a segmentation map. In \emph{Stage 2} a two class classification problem is formulated where patches from the region identified as lungs in slice $\mathbf{S_n}$ is classified into a class $C_i\in \{C_1, C_2\}$. $C_1$ corresponds to presence of nodules and $C_2$ indicates its absence.

\section{Method}

The overall workflow is divided into two stages. In \emph{Stage 1}, lung area is segmented followed by detection of nodules in \emph{Stage 2}. Overview of the method is shown in Fig.~\ref{fig:fig1}.

\noindent \textbf{Stage 1: Lung segmentation}

A modified version of SUMNet \cite{nandamuri2019sumnet} with batch-normalization is trained in \emph{Stage 1} for segmentation of lungs in 2D slices of CT. The encoder of the network is similar to VGG16 \cite{simonyan2014very} and is initialized with ImageNet pre-trained weights. The network has feature concatenation across convolutional blocks and propagation of indices across pooling layers from encoder to decoder. In addition to the structural segmentation loss, the network is trained using an additional adversarial loss similar to a generative adversarial network \cite{
goodfellow2014generative}. Fig.~\ref{fig:stage_1} shows the two steps of training involved in $Stage 1$.

\begin{figure}[h]
  \centering
  \includegraphics[width=\linewidth]{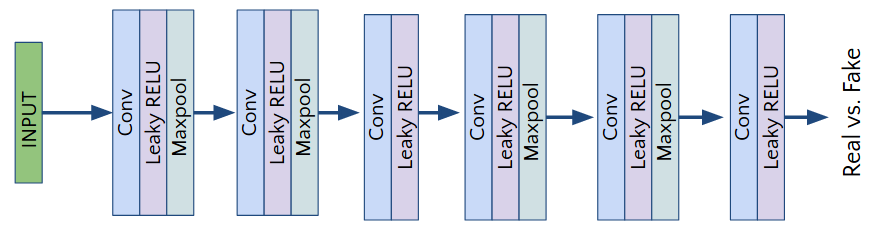}
  \caption{Architecture of discriminator}
  \label{fig:discr_arch}
\end{figure}

\begin{figure}[h]
  \centering    
    \subfigure[Training discriminator]{\includegraphics[width=0.75\linewidth]{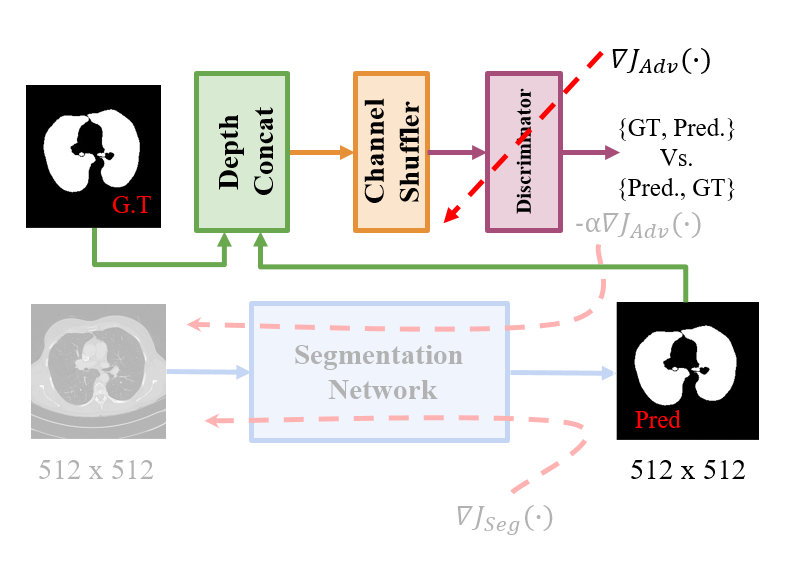}\label{fig:discr_train}}
    
    \subfigure[Training segmentation network]{\includegraphics[width=0.75\linewidth]{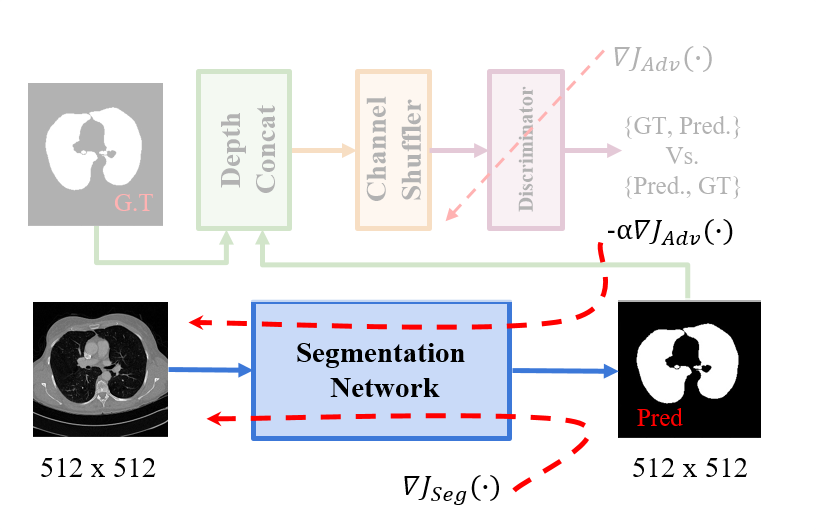}\label{fig:seg_train}}
  \caption{Figure shows training of (a) discriminator and (b) segmentation network in   \emph{Stage 1}. }
  \label{fig:stage_1}
\end{figure}

In each iteration of training, the discriminator is first trained to performs a Turing test to identify the ground truth (GT) and the segmentation map (Pred.) by presenting them together as the input to the network \cite{sathish2019adversarially}. The concatenated input tensor of size $2\times512\times512$ is shuffled along the depth to randomize the order of the two channels as shown in Fig.~\ref{fig:discr_train}. This ensures that the discriminator doesn't leverage the order of concatenation to differentiate between GT and Pred. The network is trained by minimizing the loss $J_{Adv}$, which is computed as the binary cross-entropy loss between its prediction ($\mathbf{o}_d$) and the true label ($\mathbf{t}_d$).
\begin{equation}
    J_{Adv} = \mathbf{o}_d\log(\mathbf{t}_d)+(\mathbf{1}-\mathbf{o}_d)\log(\mathbf{1}-\mathbf{t}_d)
\end{equation}
\noindent where $\mathbf{t}_d \in \{[1,0],[0,1]\}$ and depends on the order of shuffling. The architecture of discriminator is show in Fig.~\ref{fig:discr_arch}.

Next, the segmentation network is trained as shown in Fig.~\ref{fig:seg_train} using $J_{Net}$ given as,
\begin{equation}
    J_{Net} = J_{Seg} - \alpha J_{Adv}
\label{eq:total_loss}
\end{equation}  
\noindent where $J_{Seg}$ is the cross entropy loss computed between the ground truth ($GT$) and the segmentation map ($Pred$). The hyper-parameter $\alpha$, is chosen empirically to ensure similar magnitude of two losses.

\noindent \textbf{Stage 2: Detection of lung nodules}

\begin{figure}[h!]
  \centering
  \includegraphics[width=0.75\linewidth]{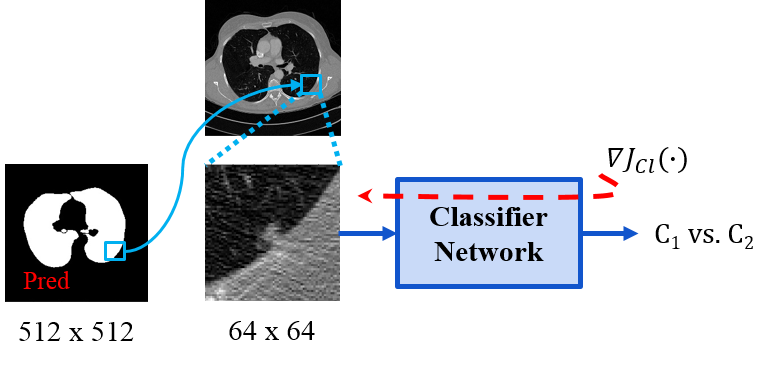}
  \caption{Figure shows the overview of \emph{Stage 2}. Class $C_1$ corresponds to presence of nodules and class $C_2$  corresponds to its absence.}
  \label{fig:classifier}
\end{figure}

The nodules are significantly smaller in size in comparison with lungs. Therefore, patches of size $64\times64$ are extracted from the region segmented as lungs in each slice in \emph{Stage 1} for further processing in \emph{Stage 2}. A LeNet \cite{lecun1998gradient} based classifier is used to detect the presence of nodules in the patches as shown in Fig.~\ref{fig:classifier}. The third convolutional layer of the standard LeNet architecture is modified to have a kernel size of $5$. An additional fully-connected layer having $256$ neurons is introduced after the last sub-sampling layer. Also, the number of neurons in last layer is changed from $10$ to $2$. Thus, by evaluating patches within the lung region, the presence of nodules in each slice of the CT volume is determined

\section{Experiments}

\noindent\textbf{Datasets:} The proposed method is experimentally validated by performing 10-fold cross-validation on the LUNA16 challenge dataset\footnote{https://luna16.grand-challenge.org/}. The dataset consists of CT volumes from 880 subjects, provided as ten subsets for 10-fold cross-validation. In each fold of the experiment, eight subsets from the dataset was used for training and one each for validation and testing. The annotations provided includes binary masks for lung segmentation and, coordinates and spherical diameter of nodules present in each slice. LIDC-IDRI dataset\footnote{http://doi.org/10.7937/K9/TCIA.2015.LO9QL9SX} from which LUNA16 is derived has  nodule annotations in the form of contours which preserves its actual shape. Therefore, we use annotations from LUNA dataset only in \emph{Stage 1}. The annotations for the nodules from the LIDC dataset is used in \emph{Stage 2} (nodule detection) to determine presence of nodules in image patches. The ground truth annotations were marked in a two-phase image annotation process performed by four experienced thoracic radiologists. Systematic sampling of slices from the CT volumes was performed to ensure equal distribution of slices with and without the presence of nodules. 

\noindent\textbf{Stage 1: Lung segmentation}

\textit{Baselines:} \label{baselines} The following baselines were used for comparison of performance of \emph{Stage 1}. \textbf{BL1}: UNet \cite{ronneberger2015u} trained using only $J_{Seg}$. \textbf{BL2}: R2UNet \cite{alom2018recurrent} trained using only $J_{Seg}$. \textbf{BL3}: SUMNet \cite{nandamuri2019sumnet} trained using only $J_{Seg}$. 

\textit{Training:} The segmentation network and the discriminators were trained using Adam optimizer \cite{kingma2014adam} for 35 epochs, with a learning rate of $10^{-4}$. The loss scaling factor $\alpha$ in Eq.~\ref{eq:total_loss} was set as $0.001$.

\noindent\textbf{Stage 2 : Detection of lung nodules}

 The classification network for nodule detection was trained using $32,594$ patches  of size $64\times64$ extracted from the lung region in the CT images, including the lung walls. Out of these patches, $16,440$ had partial or entire nodules accounting to $3,288$ in number. The model was trained for 50 epochs using Adam optimizer, with a learning rate of $10^{-4}$.

\section{Results and Discussion}

\noindent\textbf{Evaluation of lung segmentation} 

\textit{Quantitative evaluation:} The performance of the proposed method and the baselines listed in Sec.~\ref{baselines} for segmentation was evaluated using dice similarity coefficient (DSC),Area under the ROC Curve (AUC) and Hausdorff distance (HD).
\begin{equation}
    DSC = \frac{2|Pred. \cap GT|}{|Pred.|+|GT|}
\end{equation}

\begin{table}[h]
\centering
\caption{Quantitative metrics for lung segmentation (\emph{Stage 1}). }
\label{Table1}
\begin{tabular}{|c|c|c|c|}
\hline
Architecture             & DSC   & AUC    & HD     \\ \hline \hline
BL1                      & 0.979 & 0.9948 & 3.9424 \\\hline
BL2                      & 0.983 & 0.9950 & 3.8754 \\\hline
BL3                      & 0.980 & 0.9945 & 3.9596 \\\hline
\textbf{Proposed method} & 0.983 & 0.9948 & 3.8989 \\ \hline
\end{tabular}%
\end{table}

\noindent Tab.~\ref{Table1} presents the DSC, AUC Score and HD for fold 1 of the experimental setup. Due to the computational complexity of BL1 and BL2 which results in longer training duration, the performance evaluation was performed on only fold 1 of the dataset for these two methods. The mean and standard deviation of DSC across all ten folds of the experiments for BL3 and the proposed method is $0.980\pm0.0018$ and $0.984\pm0.0007$ respectively. The adversarial learning framework improves the performance consistently across all folds.


\textit{Computational complexity:} The number of trainable parameters and total floating point operations (TFLOPS) for the baselines and proposed method are shown in Tab.~\ref{Table2}. TFLOPS calculation was done for a sample input of size $512\times512$. It can be seen from Tab.~\ref{Table2} that the proposed method has lesser trainable parameters and involves significantly less number operations in comparison with the popularly used UNet (BL1) and R2UNet (BL2). Time taken to train the baselines and the corresponding model size is shown in Tab.~\ref{Table3}. All the networks were trained for 35 epochs on a GeForce GTX 1080 with 11GB RAM. Though the performance of the proposed method is comparable to that of BL2, it is computationally more efficient. The proposed framework can be trained $6\times$ faster with $1.6\times$ lesser memory requirement.

\begin{table}[!h]
\centering
\caption{Total number of trainable parameters and FLOPS during inference.}
\label{Table2}
\resizebox{\columnwidth}{!}{%
\begin{tabular}{|l|l|l|l|}
\hline
Architecture & Parameters & TFLOPs  \\ \hline \hline
BL1 & 34,525,952 (34.526M) & 261,901,254,656 (261.901G)  \\ \hline
BL2 & 39,091,328 (39.091M) & 612,076,355,584 (612.076G)  \\ \hline
BL3 & 23,863,336 (23.863M) & 153,538,789,376 (153.539G)  \\ \hline
\begin{tabular}[c]{@{}l@{}}Proposed\\ method\end{tabular} & 24,997,032 (24.997M) & 153,538,789,376 (153.539G) \\ \hline
\end{tabular}%
}
\end{table}


\begin{table}[!h]
\centering
\caption{Comparison of computational complexity based on the total training time  (35 epochs) and trained model size}
\label{Table3}
\begin{tabular}{|l|l|l|}
\hline
Architecture & Training time & Model size \\ \hline \hline
BL1 & 709m 37s & 138.2 MB \\ \hline
BL2 & 2153m 11s & 156.5 MB \\ \hline
BL3 & 326m 40s & 95.5 MB \\ \hline
\begin{tabular}[c]{@{}l@{}}Proposed\\ method\end{tabular} & 458m 17s & 95.5 MB \\ \hline
\end{tabular}%
\end{table}

\begin{figure*}[h]
    \centering
    \subfigure[Test image]{\includegraphics[width=0.18\textwidth]{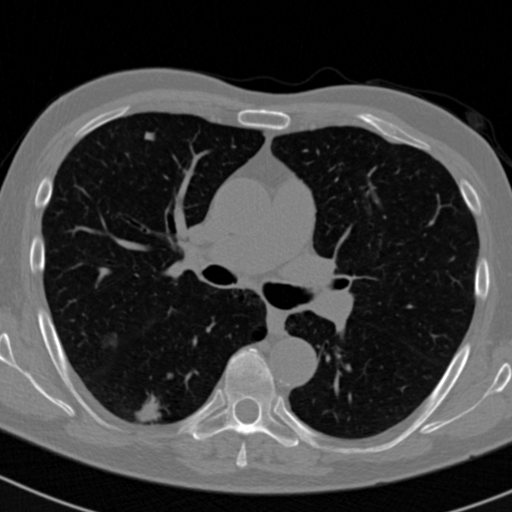}}
    \subfigure[Ground truth]{\includegraphics[width=0.18\textwidth]{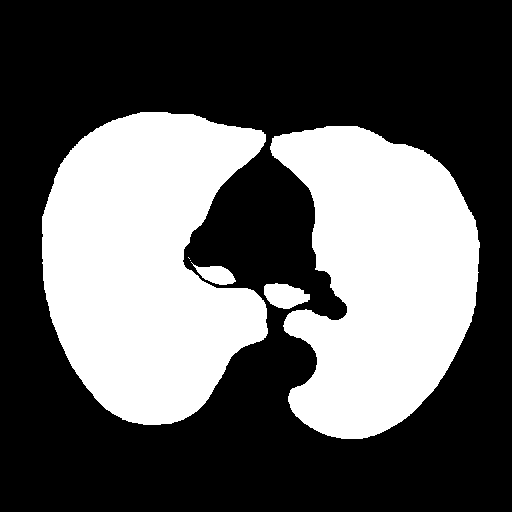}}
    \subfigure[Prediction of \emph{Stage 1}]{\includegraphics[width=0.18\textwidth]{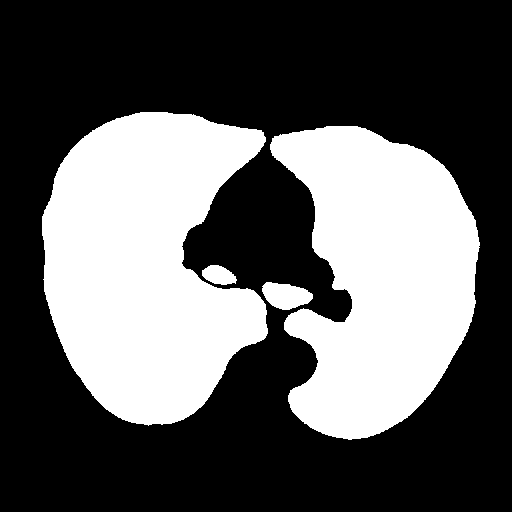}}
    \subfigure[Result of image processing ]{\includegraphics[width=0.18\textwidth]{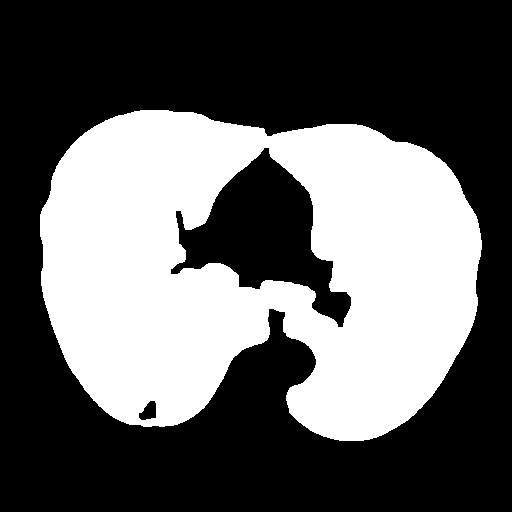}}

    \caption{Figure shows (a) a sample test image, (b) the corresponding ground truth (c) prediction of \emph{Stage 1} and (d) result of classical image processing.}
    \label{fig:qualitative}
\end{figure*}

\textit{Qualitative evaluation:} The qualitative result of the segmentation of lungs for the proposed framework on a randomly chosen test image is shown in Fig.~\ref{fig:qualitative}. The result of segmentation in the hilar region of lungs improves significantly with the proposed method.

\noindent\textbf{Evaluation of nodule detection}

\begin{figure}[h]
  \centering
  \subfigure[Image patch]{\includegraphics[height=0.33\linewidth]{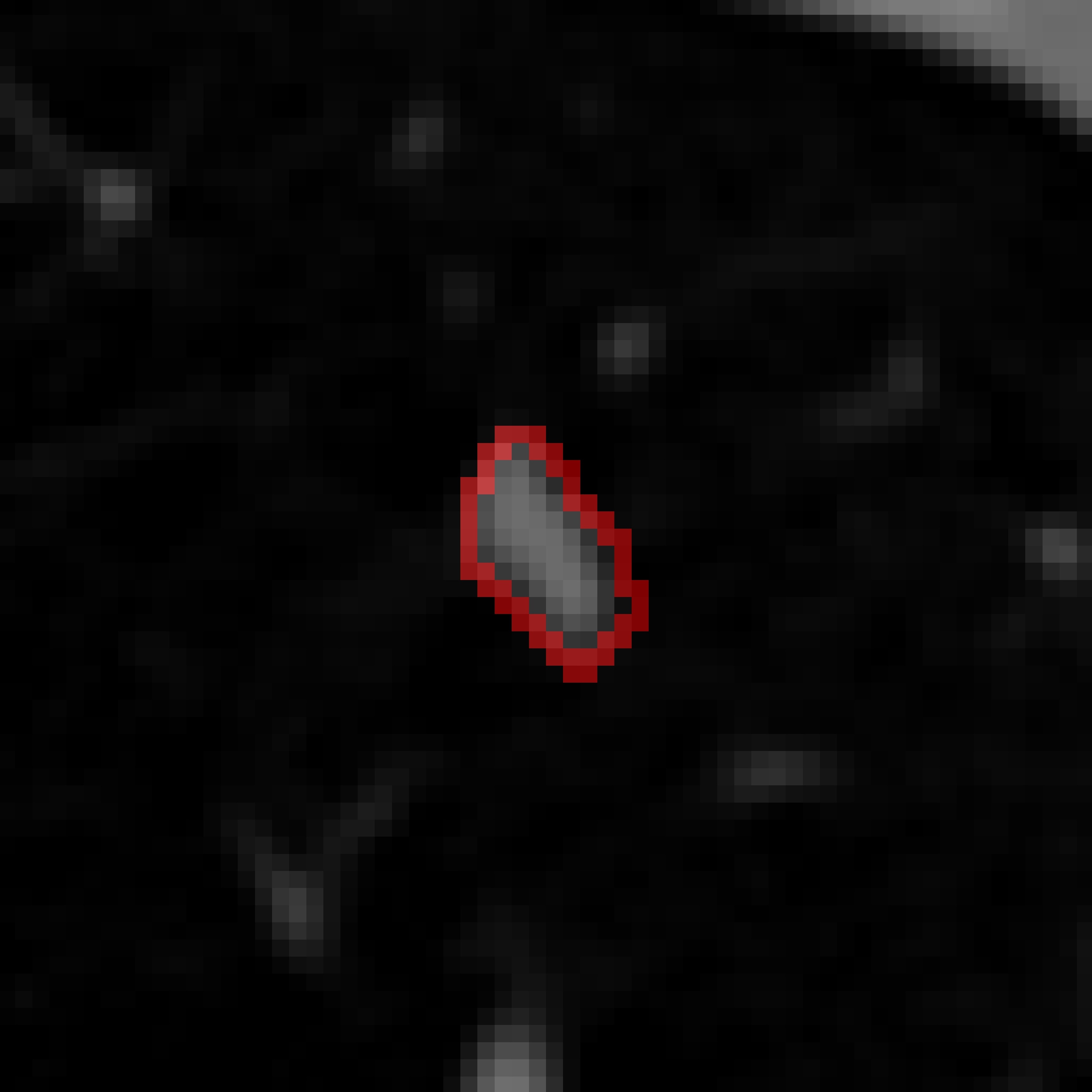}}
  \subfigure[Saliency map]{\includegraphics[height=0.33\linewidth]{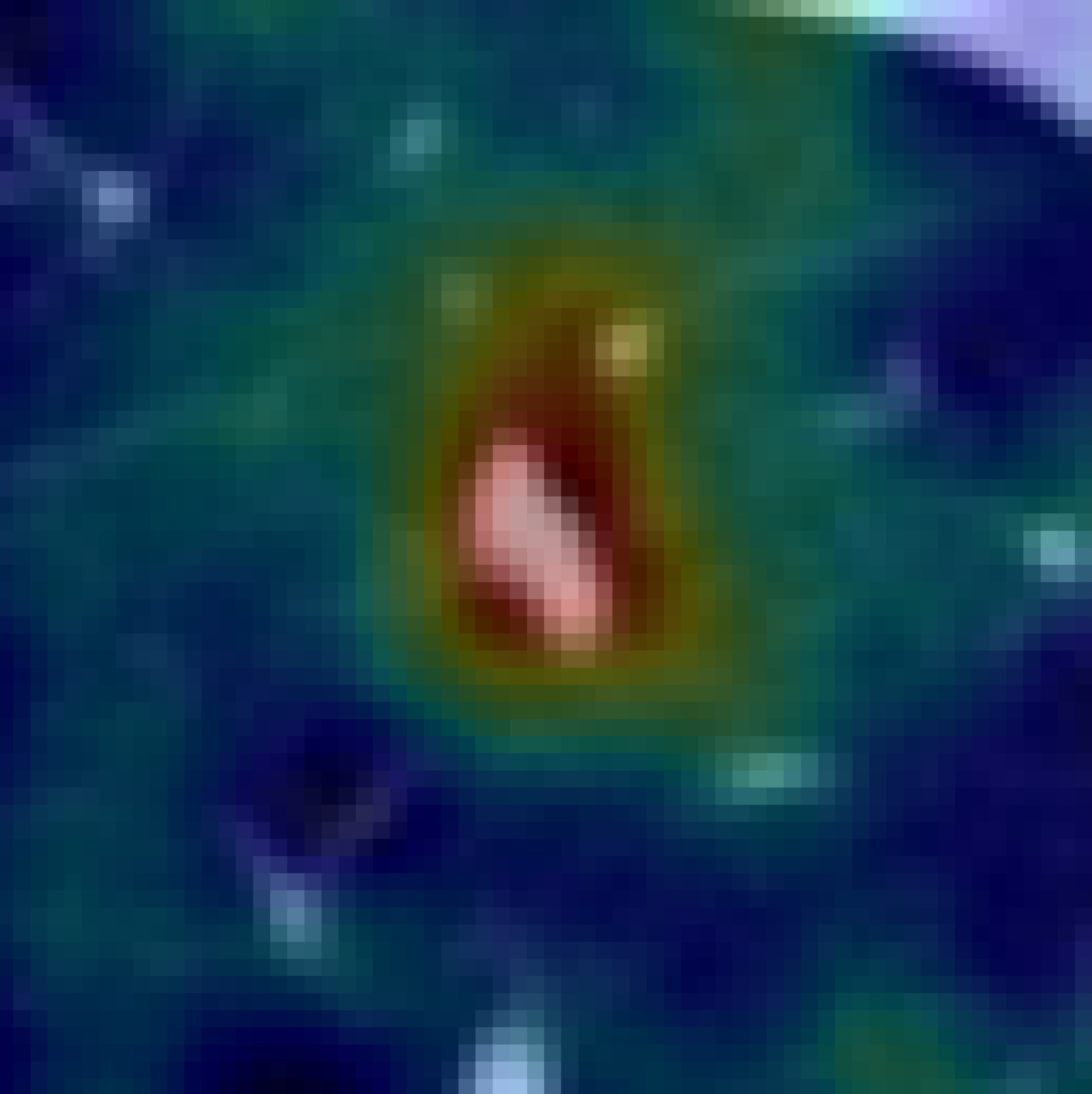}}
    
  \caption{Figure shows (a) sample image patch given as an input to the classifier network with the nodule marked in yellow (b) saliency map generated using RISE for positive prediction. In the colormap, red depicts high importance score and blue corresponds to a low score.}
  
  \label{fig:rise}
\end{figure}

\textit{Quantitative performance:}
The performance of the proposed method in detecting nodules is reported in terms of accuracy, sensitivity, and specificity of classification. Scores obtained are presented in Tab.~\ref{Table4}. Explainability of ML based methods is crucial in the field of medical image analysis. We have evaluated the trained classification network using RISE \cite{petsiuk2018rise} to obtain a saliency map which assigns an importance score to each pixel of the input  towards the model's prediction. Fig.~\ref{fig:rise} shows the saliency map for a randomly selected input patch which contains a nodule.

\begin{table}[!h]
\centering
\caption{Performance of the classification network}
\label{Table4}
\begin{tabular}{|l|l|}
\hline
Metric & Score \\ \hline \hline
Accuracy & 98.00\% \\ \hline
Sensitivity & 0.9902 \\ \hline
Specificity & 0.9688 \\ \hline
\end{tabular}%
\end{table}

\textit{Computational performance:}
Classification network is computationally lighter in comparison with the segmentation network with a total of 50,162 (50.162K) trainable parameters and 2,796,448 (2.796M) floating point operations.

\section{CONCLUSION}

In this paper, we propose a two stage framework for detection of lung nodules in CT slices. In the first stage, we segment the lung region using an adversarially trained CNN followed by detection of presence of lung nodules in image patches extracted from the lung region. We have achieved results comparable to the state-of-the-art architectures in terms of dice-coefficient score for segmentation and in terms of accuracy, specificity and sensitivity for the classification task, while being significantly less computationally expensive in terms of total number of parameters, total floating point operations, training time and model size. The additional loss from a discriminator used to train the network helps in improving the performance with very less increment in the computational complexity. Further, RISE based evaluation of explainability shows that the network attends to the relevant regions in an image while detecting the presence of a nodule.


\small
\bibliographystyle{IEEEtran}
\bibliography{refs}

\end{document}